\documentclass[a4paper]{article}

\usepackage[pages=all, color=black, position={current page.south}, placement=bottom, scale=1, opacity=1, vshift=5mm]{background}

\usepackage[margin=1in]{geometry} 

\usepackage{amsmath}
\usepackage{amsthm}
\usepackage{amssymb}

\usepackage[utf8]{inputenc}
\usepackage{hyperref}

\theoremstyle{plain}

\theoremstyle{definition}

\usepackage{graphicx, color}
\graphicspath{{fig/}}

\usepackage{algorithm, algpseudocode} 
\usepackage{mathrsfs} 
\usepackage{lipsum}
\usepackage{authblk}

\title{Computational methods for differentially expressed gene analysis from RNA-Seq: an overview}

\author[1]{Juliana Costa-Silva}
\author[2]{Douglas S. Domingues} 
\author[1]{David Menotti}
\author[3]{Mariangela Hungria}
\author[4]{Fabricio M. Lopes}

\affil[1]{Department of Informatics, Federal University of  Paraná (UFPR) - Curitiba, Paraná, Brazil}
\affil[2]{Department of Biodiversity, São Paulo State University (UNESP) - Rio Claro, São Paulo, Brazil}
\affil[3]{Department of Soil Biotecnology, Embrapa Soybean - Londrina, Paraná, Brazil}
\affil[4]{Department of Computer Science, Bioinformatics Graduate Program, Federal University of Technology - Paraná (UTFPR) - Cornélio Procópio, Paraná, Brazil}

\begin{document}

	\maketitle
Corresponding author. \href{email:fabricio@utfpr.edu.br}{fabricio@utfpr.edu.br}\\

\begin{abstract}
		The analysis of differential gene expression from RNA-Seq data has become a standard for several research areas mainly involving bioinformatics. The steps for the computational analysis of these data include many data types and file formats, and a wide variety of computational tools that can be applied alone or together as pipelines. This paper presents a review of differential expression analysis pipeline, addressing its steps and the respective objectives, the principal methods available in each step and their properties, bringing an overview in an organized way in this context. In particular, this review aims to address mainly the aspects involved in the differentially expressed gene (DEG) analysis from RNA sequencing data (RNA-Seq), considering the computational methods and its properties. In addition, a timeline of the evolution of computational methods for DEG is presented and discussed, as well as the relationships existing between the main computational tools are presented by an interaction network. A discussion on the challenges and gaps in DEG analysis is also highlighted in this review.
	\end{abstract}


\section{Introduction}
\label{sec:intro}

RNA-Seq is the standard sequencing technique to reveal the presence and quantify transcripts in a biological sample and, as a result, to allow the differential expression analysis from RNA-Seq sequences. Therefore, several methods have been proposed with different approaches and improvements in order to perform the differentially expressed gene (DEG) analysis from RNA sequencing data (RNA-Seq) \cite{Cao2020,Guo2020,Jacinto2019}. 

Especially during the last decade, RNA-seq has become an indispensable tool for DEG and splicing analysis in mRNAs \cite{Stark2019}. In this context, considering the diversity of methods proposed in the literature, some questions are interesting and deserve attention:
Which methodologies are available for DEG analysis? 
Which are the different approaches used by the methods for the definition of DEGs? 
What are the differences between the analyses performed by each method? 
And finally, how to choose a method for DEG analysis?

Some common goals are usually present in the methodologies for DEG: 
(i) to improve the accuracy of the results;
(ii) to remove biases from the analysis \cite{anders2010differential}; and 
(iii) to fill some gaps in the existing analyses \cite{trapnell2010transcript}.

However, although most methods share these goals, the computational tools developed for the DEG analysis present different approaches for evaluating their results. Therefore, defining which method to use for DEG analysis with greater precision is not a trivial question, mainly because of the large number of variables involved in this decision.

When considering the methods and their implementations (software) developed since the popularization of RNA-Seq, each variable has been faced by one or more tools in the search for increasingly efficient DEG analysis. In this sense, some studies address this context and present a review of existing approaches to analysis, their characteristics and applications \cite{Mortazavi2008, wilhelm2008dynamic, sultan2008}.

With the challenge of determining if the count of a transcript or exon is significantly different between experimental conditions, the pioneer edgeR tool for DEG analysis was developed by \cite{robinson2010edger}. Using parametric analysis and has become widely used since its creation.

Other challenge is the simultaneous transcript discovery and abundance estimation without require by prior gene annotations. Cufflinks method was developed to address these analysis issues \cite{trapnell2010transcript}.

Despite the great popularity of the computational tools initially developed \cite{hardcastle2010bayseq, trapnell2010transcript, anders2010differential}, there are still many advances needed. In this context, some tools have sought to improve aspects, such as the impact of the depth of sequencing in the identification of DEG \cite{tarazona2011differential}, the quantification of genes and isoforms with or without reference genomes \cite{Li2011}, distinct experimental conditions and their influence on these analyses \cite{Li2011}, among others.

Considering the proposed methods and software produced for DEG analysis, it is possible to observe that the choice for parametric methods \footnote{which states that the data follow a certain distribution} is recurrent and in general presents adequate results. Moreover, it is also notable that the series of steps to identify DEG needs to be computationally simplified, given the volume of pipelines developed after the RNA-Seq popularization \cite{Howe2011, trapnell2013differential, Frazee2015, Costa-Silva2017, Cao2020}.

The popularization of tools that integrate the steps of the DEG analysis (pipeline) is due in part to the fact that the analysis has several steps and, for each step, a type of file. Another challenge for the understanding and application of the methods is relies on the various options of tools for each steps. It is common to find studies of related themes that use totally different analytical tools \cite{McDermaid2019, Ren2012, Cui2015}. 

The tools for DEG analysis have some characteristics that allow their grouping. One of them, adopted in this review, is the way to treat the distribution of the expression data. The approaches that consider a certain distribution for the data analysis, i.e. that data will have a certain statistical distribution, are the parametric approaches. In contrast, some methods do not consider data distribution (data with unknown distribution), which is called non-parametric approach. One more option is to consider both or more approaches to indicate the DEGs, which are called as hybrid approaches.

In many situations, having several options can be advantageous. However, when searching for the \textit{gold standard} for a certain data analysis, it is more comfortable to visualize few options that are consolidated in the literature, as is the case of sequence alignments by BLAST \cite{lobo2008basic}.

In this context, this review reviews the main methods of DEG analysis and describes the evolution of computational methods, their properties and relationships. Moreover, a historical context is presented with the main methodologies implemented since the rise of the RNA-Seq, seeking to identify the main alternatives used to perform DEG analysis and to clarify issues about this context, such as the questions raised early in this review.

The steps involved in the differential expression analysis from RNA-Seq data will be presented in the Differential Expression Analysis section, followed by a brief history about the methodologies of differential expression analysis, discussing their main properties, similarities, differences and applications in the Methods for differential gene expression analysis section. At the end we present a discussion about the convergent and divergent points between the analyzed methodologies in the Discussion section, some conclusions and directions for DEG analysis.

\section{Differential Expression Analysis (Pipeline)}
\label{sec:pipeline}

Differential expression analysis is composed of several steps, which are presented in this section to provide an overview and the challenges involved. Naturally, the first step is to identify the main steps and the respective methods available. The second step is to choose the composition of an analysis protocol for differential expression, commonly referred to as a differential expression analysis pipeline. 

\begin{figure}[!ht]
    \center
    \includegraphics[width=1\linewidth]{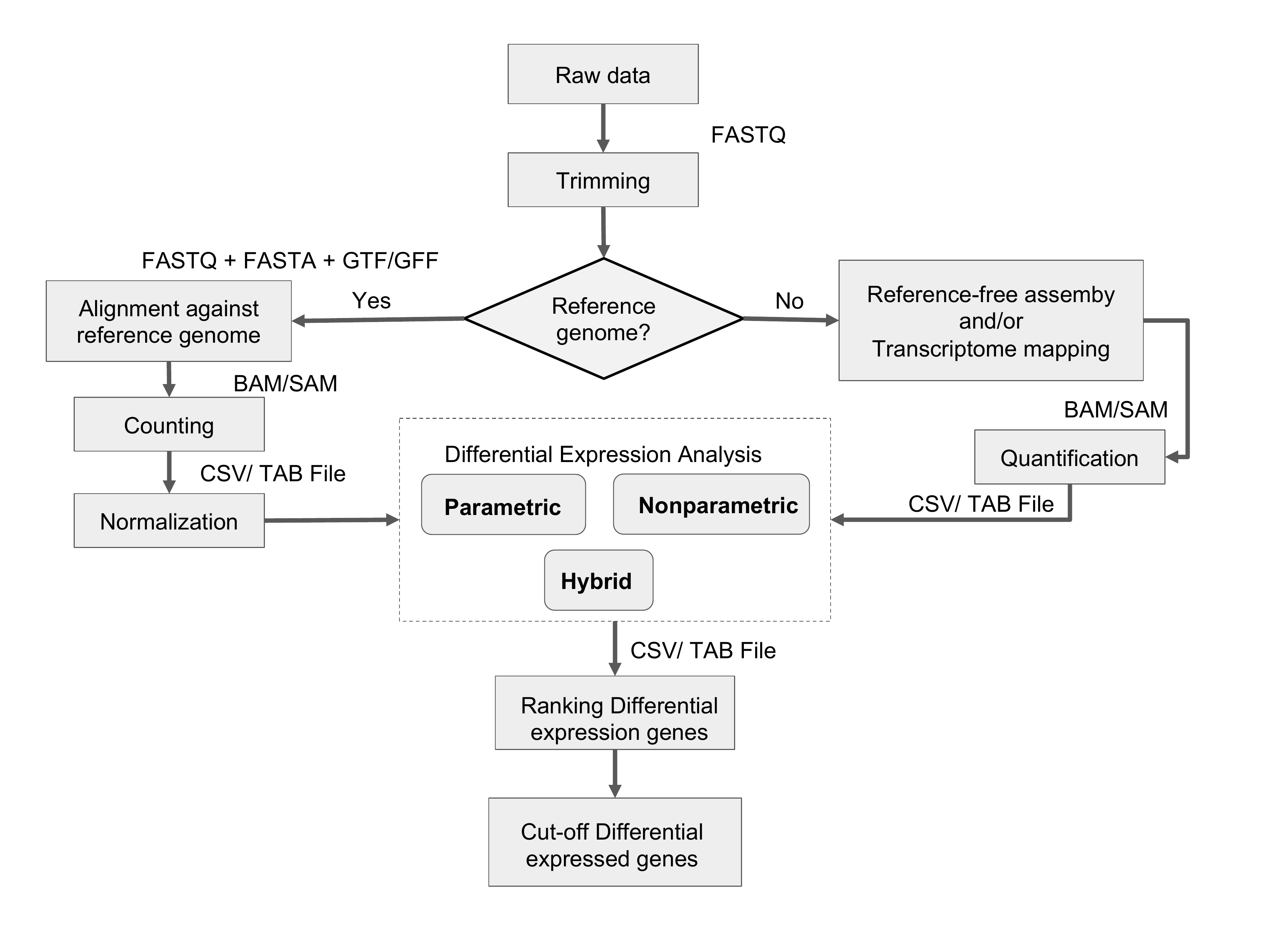}
    \caption{Overview with the main steps of differential expression analysis from RNA-Seq data.}\label{fig:pipeline}
\end{figure}

Figure \ref{fig:pipeline} presents the steps commonly used in differential expression analysis. It is possible to notice that the key steps involved are the trimming, alignment, counting, normalization and expression analysis (to define the genes/sequences with differential expression).

Considering the importance of the choice of methodologies for each of these steps, some studies sought to establish protocol for the analysis, such as \cite{Trapnell2012} and \cite{zhang2014comparative}. Instead, it is also possible to find studies that present divergences in the choices of methods in some steps of differential expression analytics \cite{Ren2012, Cui2015,McDermaid2019}.

This review aims to address mainly the aspects involved in DEGs analysis and contextualize the related methods. Although preparing sequencing libraries, reported in previous studies \cite{robinson2007moderated, Mcintyre2011, Hansen2010biases}, represent an important issue regarding its impact on the results.
Among the important points that must be considered in a differential expression analysis pipeline are (i) access to the reference genome or transcriptome; 
(ii) quality of annotations; and 
(iii) number of samples.

The next sections present some important considerations on the main steps involved in differential expression analysis.

\subsection{Quality Assessment and Trimming}
\label{subsec:quality}
The quality assessment and sequence trimming is the first step of the analysis, which is also common to other analyses involving sequencing data, such as genome and transcript assemblies \cite{Wang2009, MacManes2014}. 

Quality assessment aims to identify and remove sequences identified with low quality \cite{Li2015}. More specifically, trimming removes sequences from adapters used in sequencing. However, it is very common to find the term cleaning or trimming to refer to both steps, i.e., removal of both adapters and low quality sequences. 

Therefore, for this task, the main methodologies and implementations (software) available in the literature are presented in Table \ref{tab:clean}.

\begin{table}[!ht]
\caption{Main methodologies for the removal of adapters and low quality sequences. The methodologies are ordered chronologically considering the year of publication.}
\label{tab:clean}
\begin{tabular*}{\columnwidth}{@{\extracolsep\fill}lcccc@{\extracolsep \fill}}
Name              & Adapter & Quality & Year & Reference \\ 
Btrim             & Yes & Yes & 2011 & \cite{Kong2011} \\ 
CutAdapt          & Yes & Yes & 2011 & \cite{EJ200} \\ 
Trimmomatic       & Yes & Yes & 2014 & \cite{Bolger2014} \\ 
AdapterRemoval v2 & Yes & Yes & 2016 & \cite{Schubert2016} \\ 
Atropos           & Yes & Yes & 2017 & \cite{Didion} \\ 
fastp             & Yes & Yes & 2018 & \cite{Chen2018}\\
\end{tabular*}
\end{table}

The sequence trimming process evaluates the quality of the reads\footnote{reads: small DNA sequences from sequencing} obtained in sequencing. 

Two characteristics are evaluated, the presence of adapter sequences (used in sample preparation) and the quality of each base pair read by the sequencer. As a result, the reads that get a certain score within the scale defined by the \cite{Williams2016} sequencing technique are selected.

In this process, each base pair is evaluated from the quality score informed by the sequencer. It is possible to choose the cut score, besides defining the reads that should be kept or discarded.

Considering the identification of the quality of the bases, the quality information in the FASTQ \cite{Cock2009} file is used. In contrast, the removal of sequences identified as adapters is performed by similarity search \cite{Bolger2014} in the sequences.

\subsection{Alignment} 
\label{subsec:mapping}

After trimming the sequences, the mapping and counting of mapped reads occurs. In this process, the aim is to identify how many reads are aligned to a region of the genome. The result is a read count table aligned to each gene. Among the difficulties in the mapping process, the processing time and the computational capacity used are the major challenges.

Table \ref{tab:mapp} presents the main mapping methodologies, where it can be observed that the Burrows-Wheeler Transformation algorithm \cite{burrows1994} is recurrently applied by the methodologies. Its popularity is mainly because the transformation algorithm is computationally fast, allowing the mapping method can be executed in common computers (desktop).

\begin{table}[!hb]
\caption{The principal sequence mapping methods.}
\label{tab:mapp}
\begin{tabular*}{\columnwidth}{@{\extracolsep\fill}llll@{\extracolsep\fill}}
Name    & Algorithm & Year & Reference \\ 
BWA     & Burrows–Wheeler Transform & 2009 & \cite{Li2009} \\ 
RUM     & Burrows–Wheeler Transform & 2011 & \cite{Grant2011} \\
Bowtie2 & Burrows–Wheeler Transform & 2012 & \cite{langmead2012fast} \\
BWBBLE  & Burrows–Wheeler Transform & 2013 & \cite{Huang2013} \\ 
STAR    & Maximal Mappable Prefix (MMP) & 2013 & \cite{Dobin2013} \\
Tophat2 & Burrows–Wheeler Transform & 2013 & \cite{Kim2013} \\ 
HISAT2  & Graph Based & 2019 & \cite{Kim2015,Kim2019}\\ 

\end{tabular*}
\end{table}

The principal sequence mapping methodologies are briefly described in the following.
\begin{itemize}
   \item \textbf{BWA} \cite{Li2009}: is based on the backward search associated with the Burrows-Wheeler transform. This method is used to efficiently align short reads to large reference sequences such as the human genome. BWA, allows gaps and mismatches. In terms of memory optimization and search technique used, it performs a similar strategy adopted by the Bowtie method \cite{langmead2009ultrafast};
   
   \item \textbf{RUM} \cite{Grant2011}: is based on an aggregation of methods, in which reads are mapped against the genome and transcriptome using the Bowtie tool. The reads not mapped by Bowtie are aligned to the genome with the BLAT tool \cite{Kent2002a}. The result of the mappings is presented in SAM format;
   
   \item \textbf{Bowtie 2} \cite{langmead2012fast}: describes the application of the Burrows-Wheeler transform. Thus, for each read the method performs four main steps (i) the extraction of  ``seeds’’ (sequence snippets) from the read and their reverse complement; (ii) the alignment of the seeds to the genome (reference), producing Burrows-Wheeler alignment bands; (iii) the selection of the bands randomly and repeatedly (weighted by priority), applying the displacement of each selected lane on the reference genome, using a method to compress the suffix matrix, and still effectively support search for arbitrary patterns, called FM index \cite{Ferragina2000}, and applying the ``walk-left’’ strategy of the FM index; and (iv) the resolution of similar alignments, observing the edges. More details about this algorithm and its operation of Bowtie 2 are given in \cite{langmead2012fast};
   
    \item \textbf{BWBBLE} \cite{Huang2013}: is based on mapping using multiple genomes as reference, proposing the concept of a linear reference multi-genome. This concept incorporates the catalogue of all known gene variants with a reference genome (e.g. SNPs, insertions, deletions and inversions), and uses a read alignment algorithm based on the Burrows-Wheeler transform;
    
    \item \textbf{STAR} \cite{Dobin2013}: was proposed to specifically address many of the challenges of mapping RNA-Seq data, such as junction detection and characterization, and mapping sequences derived from non-contiguous genomic regions. In addition, it uses a novel strategy for junction alignments. The alignment comprises two major steps: seed search step and clustering step. The main idea behind the STAR seed search step is the sequential search for a Maximum Mappable Prefix (MMP). In the clustering step, STAR builds alignments of the entire read sequence by joining all the seeds that were aligned to the genome in the first step.
    
    \item \textbf{Tophat 2} \cite{Kim2013}: directs attention to the problem of multiple alignments injunction reads. It uses the Bowtie 2 tool as a dependency. In the mapping step, reads aligned to more than one exon are treated as unmapped. These reads are fragmented and aligned to the genome. Tophat 2 considers that the alignment distance between fragments may indicate possible splice regions. The genomic sequences around these junction sites are concatenated, and the resulting spliced sequences are treated as a set of potential transcription fragments. Any reads not mapped in the previous stages (or poorly mapped) are then realigned with Bowtie2 against this new transcript.
    
    \item \textbf{HISAT2} \cite{Kim2019}: this method has a graph-based search strategy as its main characteristic. HISAT2 starts the alignment process by generating a linear graph of the reference genome. Then it adds mutations, insertions and deletions as alternative paths of the graph. The authors claim that graph representations are more efficient in terms of memory utilization and/or alignment speed compared to linear reference representation of genomes and alleles.

\end{itemize}

Considering the summarized presentation of the alignment methodologies, it is possible to highlight that each method presents a main strategy and is concerned with some specific alignment problems, which must be taken into consideration during the selection of the differential expression analysis pipeline. 
The evolution of methods is also noted, and the tendency to use several genomes as reference, as observed in HISAT2 \cite{Kim2019}, presented as the successor of Tophat. Further information on mapping techniques can be got through studies as \cite{Canzar2017}.

As a result of mapping, the tools produce mapping files in formats such as SAM (Sequence Alignment/Map format) and/or BAM (Binary Alignment/Map format) \cite{Li2009}. 
The resulting file contains all the mappings of a read and information such as alignment position, and alignment score identified as MAPQ (acronym for MAPping Quality), which is presented in Phred (Phred-scaled) scale. 
Phred (Q) scale \cite{Ewing1998} is a quality indicator based on the probability of error in the alignment of a read at a reference position.

The way of considering these mappings allows variations. It is possible to consider only mappings with Phred values above a threshold, or reads that obtained unique mapping (in only one region of the reference genome). To support this task, counting tools are used, which process files in SAM and/or BAM format, and produce a table with genes and amount of mapped reads. In Section Counting some counting tools and their main properties are presented.


\subsection{Pseudo Alignment}
\label{subsec:pseudo}

Besides the Burrows-Wheeler transform and graph-based alignments, there are strategies that prioritize the balance between computational performance and the results produced. In this context, the tools that use the pseudo alignment strategy are applied.

It is important to note that most of the mapping tools described earlier could be applied to analyses in which the reference genome and its respective annotation are available. When an assembled genome is not available and there is a need for mapping without reference, the transcriptome is assembled and expression is estimated based on this assembly. In general, there are two types of approaches to transcriptome assembly; (i) genome-guided (or genome-based) assembly; and (ii) \textit{de novo} assembly.

Some tools have been developed to generate transcript identification, such as Trinity \cite{grabherr2011full} and Oases \cite{Schulz2012}.
There are also strategies to estimate expression levels from data that lack a reference genome, such as RSEM \cite{Li2011} and eXpress \cite{Roberts2013}.
Some strategies perform both steps, transcript identification and estimation of expression levels, such as Scripture \cite{Guttman2010}, Cufflinks \cite{trapnell2010transcript} and StringTie \cite{Pertea2016}, Salmon \cite{Patro2017SalmonExpression} and Kallisto \cite{bray2016near}. 
Due to this dual functionality, these methods are also discussed in the section on differential expression analysis methods and included in the Supplementary Material 1.

One important method that uses pseudo alignment and transcript quantification is Salmon \cite{Patro2017SalmonExpression}. Salmon uses the quasi-mapping strategy \cite{Srivastava2016}, requires a set of transcripts as a reference (which can be a reference assembly or \textit{de-novo}) or only the reads, to quantify the transcripts. This strategy comprises three steps (i) a simplified mapping model, (ii) a phase that estimates initial expression levels and model parameters, and (iii) a phase that refines the expression estimates.
This inference procedure allows Salmon to build a probabilistic model based on the sequencing data, which includes more information, and then improves the conditional probability that a fragment is part of a transcript.

In pseudo alignment strategies, it is important to point out that the tools for transcript identification and quantification will report normalized quantification values as output and not counts of mapped reads. Therefore, the choice of the tool for differential expression analysis should consider the type of data input expected (count or normalized values).

\subsection{Counting}
\label{sec:counting}


The count of mapped reads is the step where it will be identified how many reads were mapped in each genomic region (reference). This step does not define which genes are differentially expressed; however, it represents the basis for the following steps of the analysis. Because, for each sequencing file presented to the mapper, a count of reads mapped to a particular gene will be produced.
Consequently, there is need of an annotation file of the reference genome. 

The annotation file is usually in GFF (General Feature Format), which consists of one row per feature and, each row presents 9 columns of data.
The columns of the file are separated by tabs and are arranged in the following order
\textless seqname \textgreater, \textless source\textgreater, \textless feature \textgreater, \textless start \textgreater,  \textless end \textgreater, \textless score \textgreater, \textless strand \textgreater, \textless frame \textgreater $[attributes]$ $[comments]$ ($<>$ mandatory fields and $[]$ optional fields) \cite{Zhang2016}. The GTF (General Transfer Format) is identical to the GFF in its version 2. Files have the same format and fields, but with different extensions.
The description of the mandatory fields is as follows:
\begin{itemize}
    \item seqname: sequence name, generated by a software or can be access from a public database such as Genbank \cite{Sayers2019} or EMBL \cite{Hubbard2009};
    \item source: name of the software that generated the feature or data source; 
    \item feature: name of the feature, e.g. gene, exon, transcript; 
    \item start: initial position of the feature, the value must start at 1;  
    \item end: end position of the feature;
    \item score: a floating point value;
    \item strand: value set as (reverse) or + (forward);   
    \item frame: This value is indicated by one of the values: 0, 1, 2 or ``.'', 0 indicates that the mapped region is within the frame (no extra bases); 1 indicates an extra base, i.e. second base of the region indicates first base of the codon; 2 indicates that the third base of the region is the first base of the codon; If the frame is not relevant, the indication will be ``.''; 
    \item attribute: a semi-colon separated list indicating additional information on each feature.
\end{itemize}

GFF is a widely used text file format for storing genome annotations, describing sequence-based annotations. In addition, GFF files present genome features in a tab-delimited, single-feature-per-line table, making it ideal for use with multiple \cite{Rastogi2014} data analysis pipelines.
The GFF file is used to translate the alignment information, which presents the following information: reads A and B were mapped on the X-chromosome, between base pairs i and j.
Therefore, it indicates to which gene of the X chromosome the base pairs i and j refer.
On the other hand, the annotation file will indicate that gene Z has part or all of its sequence between positions i and j of the X chromosome, therefore reads A and B will be considered in the mapping count of gene Z.

Before performing the count it is necessary to consider the options for alignments, such as (i) read fully aligned to a gene; (ii) read partially aligned to a gene; (iii) read aligned to a junction (intron and exon); (iv) read aligned to a junction of exons (no alignment with intron); (v) read partially aligned to two genes; and (vi) read aligned to two genes. For this task there are some methods that can be used associated or isolated, such as the HTSeq-count \cite{anders2014} which is part of the HTSeq framework, the BEDTools \cite{Quinlan2010} toolkit and the featureCounts \cite{Liao2014}. Implemented in R language, the Rsubread \cite{Liao2019} method has functionality for alignment, quantification and analysis of RNA-Seq data and can also be a counting option.

The choice of method and how to consider mappings in the count should be made based on the dataset and its properties. For eventual situations where little prior knowledge is available, it is recommended to compare the count in the most restrictive mode and in the most permissive mode of each tool to define an adequate parameterisation.

\subsection{Normalization}
\label{subsec:normalization}

This step of differential expression analysis aims to define which variations in mapping count will be considered as differential expression.
If a gene has more mapped reads, it does not mean that it is differentially expressed, because as each gene has an extension of base pairs, a smaller sequence in base pair sizes may have a proportionally smaller quantity of aligned reads. In this context, normalization is one of the basic steps for the differential expression analysis.

The RPKM (Reads per Kilobase per Million) method was proposed in 2008 to generate accurate quantification of gene expression from RNA-Seq \cite{Mortazavi2008} data. This method normalizes the expression of RNA-Seq data using as a basis the total transcript size and the number of \textit{reads} sequenced. In this way, RPKM allows small genes or transcripts not to be penalized compared to larger sequences.

The FPKM normalisation approach is analogous to RPKM, but supports one, two or more sequences from the same molecular source \cite{trapnell2010transcript}. FPKM considers fragments and reads. In paired-end experiments, forward and reverse reads of the same sequence are considered as a fragment. 

In 2012 the TPM (transcripts per million) normalization approach was presented as a modification of the RPKM approach and aiming to remove RPKM bias \cite{wagner2012tpm}.

With normalized count data, it is possible to identify coherently the expression variation evaluated under different conditions. For these analyses there are several methods that use as strategy the identification of expression variations assuming that these data follow a certain statistical distribution or not.

A review of the main methods, their characteristics and applications are presented in the next section, the main focus of this review.

\section{Methods for differentially expressed gene analysis} 
\label{sec:expression}

While the overview on differential expression and its steps is presented, this review focuses on the presentation and discussion of differential expression identification/inference tools and their properties.

In the identification of DEG the aim is to infer which genes have decreased or increased transcriptional activity in certain experimental assays.
Thus, the methods for identifying DEGs consider the quantification of RNA transcription.
There are several ways to quantify transcripts, as presented briefly in the Pseudo Alignment section and to define differential expression. 
This review focuses primarily on computational tools that identify differential gene expression.

To contextualize the differential expression analysis is necessary to go back to 1991, when the identification of the transcriptional profile in mammals was proposed \cite{adams1991complementary} using the EST (Expressed Sequence Tag) technique, based on the partial sequencing of cloned cDNAs to evaluate expression. 
Some years later the SAGE (Serial Analysis of Gene Expression) method \cite{velculescu1995sage} was proposed and, in parallel, publications using the Microarray technique emerged \cite{schena1995microarray}, which became for years the most popular choice among transcriptional pattern studies.

In 2006 the first study with RNA-Seq (high-throughput platform mRNA sequencing) data \cite{bainbridge2006analysis} was published, using Roche’s 454 technology. This sequencing technique generates large numbers of short reads, in platforms that perform this type of sequencing are called high-throughput sequencers.

\begin{figure*}[!ht]
    \center
    \includegraphics[width=\linewidth]{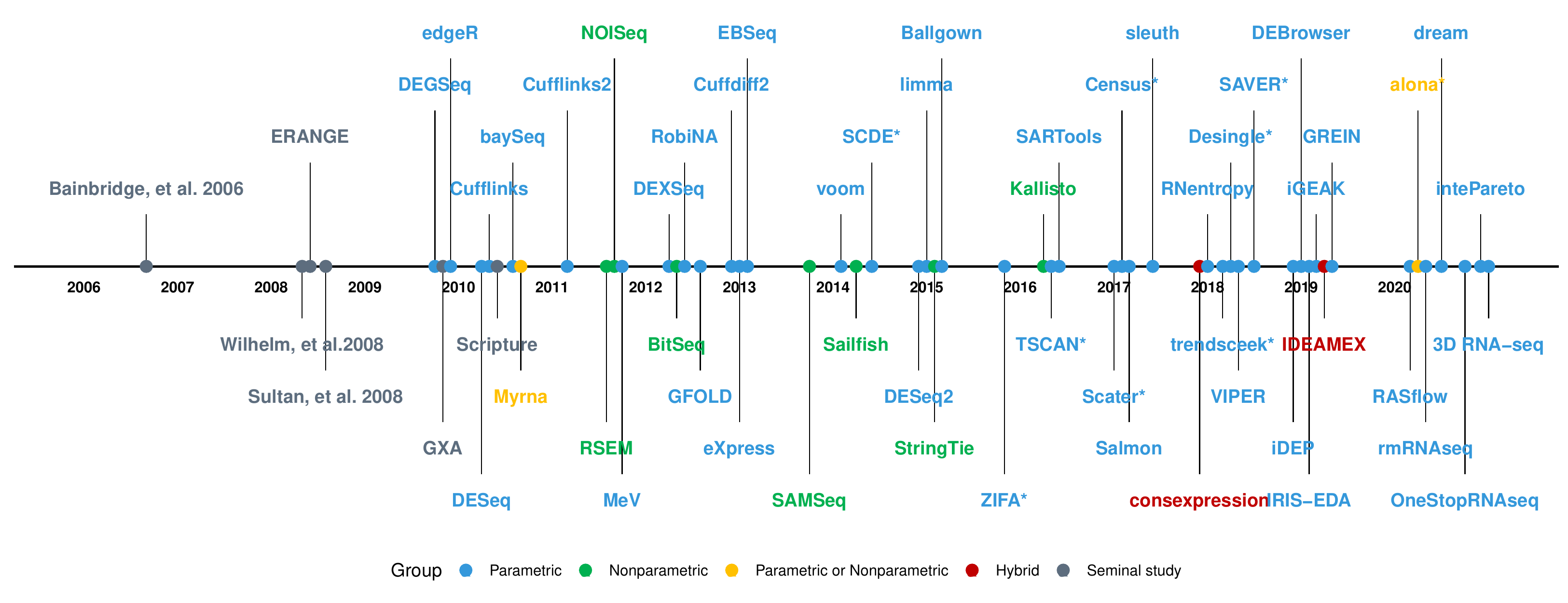}
    \caption{Timeline with the main methodologies and computational tools for DEG analysis. In blue are shown the computational tools that use parametric methods to indicate differentially expressed genes. In green the non-parametric tools and in yellow the tools that allow the use of parametric or non-parametric methods. In red are the tools considered hybrid, for using parametric and non-parametric methods together in the indication of DEG. The items that are identified in grey are pioneer publications and/or publications that boosted the analysis. The distribution of methods in the timeline considers the month and year of publication. The items that contain * in the name indicate methods developed in the context of single-cell sequencing analysis.}
    \label{fig:timeline}
\end{figure*}

Initially, aiming at presenting an overview and the relations between the main methodologies and the computational tools available in literature, the most cited ones since the beginning of RNA-Seq popularization in 2009 were identified. 
These methodologies were organized in temporal form, generating a timeline, which is presented in Figure \ref{fig:timeline}.

The confirmation of the popularity of the RNA-Seq technology occurred in 2008, with the encouragement of a trio of scientific  \cite{Mortazavi2008, sultan2008, wilhelm2008dynamic}, bringing novel approaches for the analyses. 
These studies did not specifically generate computational tools, however they pave a way for the expression analysis from RNA-Seq data. 

For this reason, these studies are identified in Figure \ref{fig:timeline} in gray as a group of seminal studies.

The initial studies sought to build a consensus on the methodologies for DEG analysis.
In this context, the study of \cite{Mortazavi2008} proposed the ERANGE (Enhanced Read Analysis of Gene Expression) software, while in the study of \cite{sultan2008} the expression data were only normalized.
The study of \cite{wilhelm2008dynamic} performed a correlation with hybridization data.

Among the most popular technologies for generating RNA-Seq data are Illumina Genome Analyzer and HiSeq \cite{mcgettigan2013transcriptomics}, which enable the production of single or paired-end reads.
In this way RNA-Seq also produces quality mappings, accurate identification of alternative splicing, transcript reconstruction, among other types of studies. Regarding its predecessor, the Microarray, RNA-seq allows the study of new transcripts, offers higher resolution, better detection range and less technical variability. These factors have led to a major expansion of RNA-Seq, becoming the first choice in transcriptome analysis for many research groups \cite{Corchete2020}.

With the popularization of the RNA-Seq technique, computational tools (softwares) and proposals of new methodologies for DEG analysis emerged. Since the proposal of the ERANGE tool proposed by \cite{Mortazavi2008}, several other tools have been and continue to be proposed in the literature, as presented in Figure~\ref{fig:timeline}.

To evaluate gene expression data generated with RNA-Seq, usually after mapping and counting mapped reads, the starting point is the decision on the type of tool to be used to identify differentially expressed genes.
In this review, the methodologies were considered regarding the statistical distribution applied in the differential expression analysis, proposing a division into three groups: parametric, non-parametric and hybrid.
Only computational tools for DEG analysis, implemented and made available in software, regardless of the programming language or form of access, were considered. 
Methodologies without computational tools were not considered, since this is an essential criterion for application in real problems.

All computational tools that adopt or describe the use of some parametric statistical distribution for the inference of DEGs were considered as parametric, as well as the tools that use totally or partially this class of statistical distribution.
The tools without use of parametric distributions in their analysis and/or, that do not present any a priori statement about the data distribution for inference of DEG were considered as non-parametric.
In this review, the computational tools described as hybrid use parametric methods associated with non-parametric methods for DEG inference.

The explanation considering parametric, non-parametric and hybrid was adopted in this review to present the main methodologies in the literature in order to organize and contextualize them. Considering this scenario, the next sections will describe the respective particularities of parametric, non-parametric and hybrid methods.

\subsection{Parametric methods}
\label{subsec:parametric}

Parametric methodologies are those which start from the premise that data present a certain distribution. When using these tools, it is considered that the input data are distributed according to the statistical distribution adopted by the method, such as negative binomial, Poisson or Gaussian.
This strategy is adopted by the first computational tools developed for DEG analysis \cite{hardcastle2010bayseq, robinson2010edger, anders2010differential}. 

Poisson distribution is adopted with some frequency for the representation of RNA-Seq data by computational tools \cite{Wang2010, Langmead2010, Feng2012}. 
Methods that use parametric analysis represent the vast majority of the tools developed and made available for use since the popularization of RNA-Seq, as presented in Figure \ref{fig:timeline}.

Considering the choice of parametric methods for DEG analysis, some considerations are essential. 
The parametric methods assume that the expression data is distributed according to a statistical distribution. 
Therefore, the identification of differentially expressed genes in this context can be defined as the genes that are at the extremities of the distribution chosen, according to the experiment and sampling using a statistical significance value as the $p$-value.

Among the most widely used distributions for DEG analysis are the Negative Binomial \cite{robinson2007moderated, hardcastle2010bayseq, robinson2010edger, anders2010differential}, Poisson \cite{marioni2008normalization, bullard2010evaluation, Wang2010, Langmead2010, Feng2012} and Gaussian \cite{Bloom2009, ritchie2015limma, Pimentel2017} distributions.

The Poisson distribution is characterised by its suitability in application to technical replicate data \cite{marioni2008normalization}. 
On the other hand, data from biological replicates have higher variance, and for this reason are best represented by a negative binomial distribution \cite{li2013finding, Howe2011, trapnell2013differential, leng2013ebseq, Varet2016}.
Gaussian distribution or normal distribution is a bell-shaped curve, and it is assumed that during any measurement values will follow a normal distribution with an equal number of measurements above and below the mean value \cite{Dasgupta2014}. The Gaussian or normal distribution is used in simulation data \cite{Pimentel2017}, parameters estimation \cite{ritchie2015limma} and in expression analysis with Microarray data \cite{Bloom2009}. 

These and many other characteristics of parametric methods made them a suitable model to be followed, which generated a large volume of tools that use parametric methods of analysis, as visualised in Figure \ref{fig:timeline}, in which many of the initially proposed methods were the basis for most of the recently proposed methods.

\subsection{Non parametric methods}
\label{subsec:nonparametric}

The nonparametric methods for DEG analysis arise in a context of innovation, with the need for solutions to the analysis of experiments with few replicates, in which the estimation of variance with precision becomes difficult.
By observing the distribution of the groups of methods in Figure \ref{fig:timeline} it is possible to notice that, between 2010 and 2013, were presented the main nonparametric computational tools for DEG analysis \cite{tarazona2011differential, Li2011, Glaus2012, li2013finding}.

Non-parametric methods include inference, non-parametric descriptive statistics, statistical models, and statistical tests. 
These methods do not determine a data distribution model a priori. 
The structure of the models is defined based on the distribution of the data, commonly known as data-driven. 
The non-parametric expression, associated with a tool for DEG analysis, shows that the number and nature of the parameters are adjusted according to the distribution of the data\cite{Penfold2012}.

Among the main tools identified in this review, NOIseq is a method that assesses differential gene expression between groups through the relationship between expression change and absolute expression differences \cite{tarazona2011differential, tarazona2015data}. NOIseq uses mapped, corrected and normalised count reads, models the noise distribution by contrasting the logarithm of expression change and absolute expression differences between groups. NOIseq defines a gene as differentially expressed between groups if the corresponding logarithm of expression change and absolute expression difference values have a high probability of being higher than the \cite{Li2019} noise values.

Another tool for non-parametric DEG analysis is SAMseq \cite{li2013finding}.
For comparisons between groups, SAMseq uses the Wilcoxon two-sample classification statistics.
On the other hand, SAMseq considers the different depths through a re-sampling process in differential data analysis.
In the Wilcoxon and FDR classification statistics, the null distribution is estimated using the \cite{Li2019} mutation method.

The RSEM \cite{Li2011} method uses the pseudo-alignment assumption and is not a tool focused solely on expression analysis. RSEM defines a probabilistic model for RNA-Seq data and calculates maximum likelihood estimates of isoform expression levels using the Expectation-Maximization algorithm \cite{Nicolae2011}.

\subsection{Hybrids}
\label{subsec:hybrids}

For this review, we considered as hybrid methods those approaches that associate parametric and non-parametric methodologies for the identification of differentially expressed genes.

The approach identified as hybrid by this review was the consexpression \cite{Costa-Silva2017} and IDEAMEX \cite{Jacinto2019}.
The consexpression approach is a pipeline for expression analysis that adopts the identification of DEGs from the joint analysis of nine tools. 
Among them, two are non-parametric: NOISeq \cite{tarazona2011differential} and, SAMSeq \cite{li2013finding}, and seven parametric \cite{hardcastle2010bayseq, anders2010differential, robinson2010edger, leng2013ebseq, Love2014, ritchie2015limma, Pimentel2017}. 
Genes so indicated by the consensus of five or more tools are considered differentially expressed. In addition, the tools for the option of parametric or nonparametric analyses show in Figure \ref{fig:timeline} use one method according to the user’s choice in an isolated manner.

In IDEAMEX, DEGs are indicated by NOISeq \cite{tarazona2011differential}, limma-Voom \cite{ritchie2015limma}, DESeq2 \cite{Love2014} and edgeR \cite{robinson2010edger},and are generated reports for each method. In result integration: the obtained results are reported using different graphical outputs, such as correlograms, heatmaps, Venn diagrams and lists \cite{Jacinto2019}. 


\section{Discussion}
\label{sec:discussion}

This review presents a temporal overview of the computational tools developed for differential gene expression analysis. The processing of this analysis is developed in several steps, as briefly described in the Methods for DEG analysis section. 
Then, based on the main methods identified in the literature and addressed in this review, their properties and applications are presented.

In the context of the trimming step, the study presented by \cite{Williams2016}, the authors state that by applying more aggressive parameters, at the sequence trimming stage, over ten percent (10\%) of the genes had significant changes in estimated expression levels. In another study, it is suggested that a softer cutoff, specifically of those nucleotides whose Phred quality rate score $<2$ or $<5$, is ideal for most studies, although a very aggressive quality cut is commonly adopted \cite{MacManes2014}.
Still regarding the quality, several studies report that by applying more aggressive and commonly used parameters such as Phred quality rate $>20$ and read size $>50$bp, no significant differences in results are found \cite{Corchete2020, Liao2020}, suggesting that a soft cut or even no cut at all results in the most biologically accurate gene expression estimates.

In \cite{Williams2016}, the authors also indicate that most expression changes could be mitigated by imposing a minimum length filter after the cutoff, suggesting that differential gene expression may be driven primarily by spurious mapping of short reads.

\begin{figure}[!ht]
    \center
    \includegraphics[width=\linewidth]{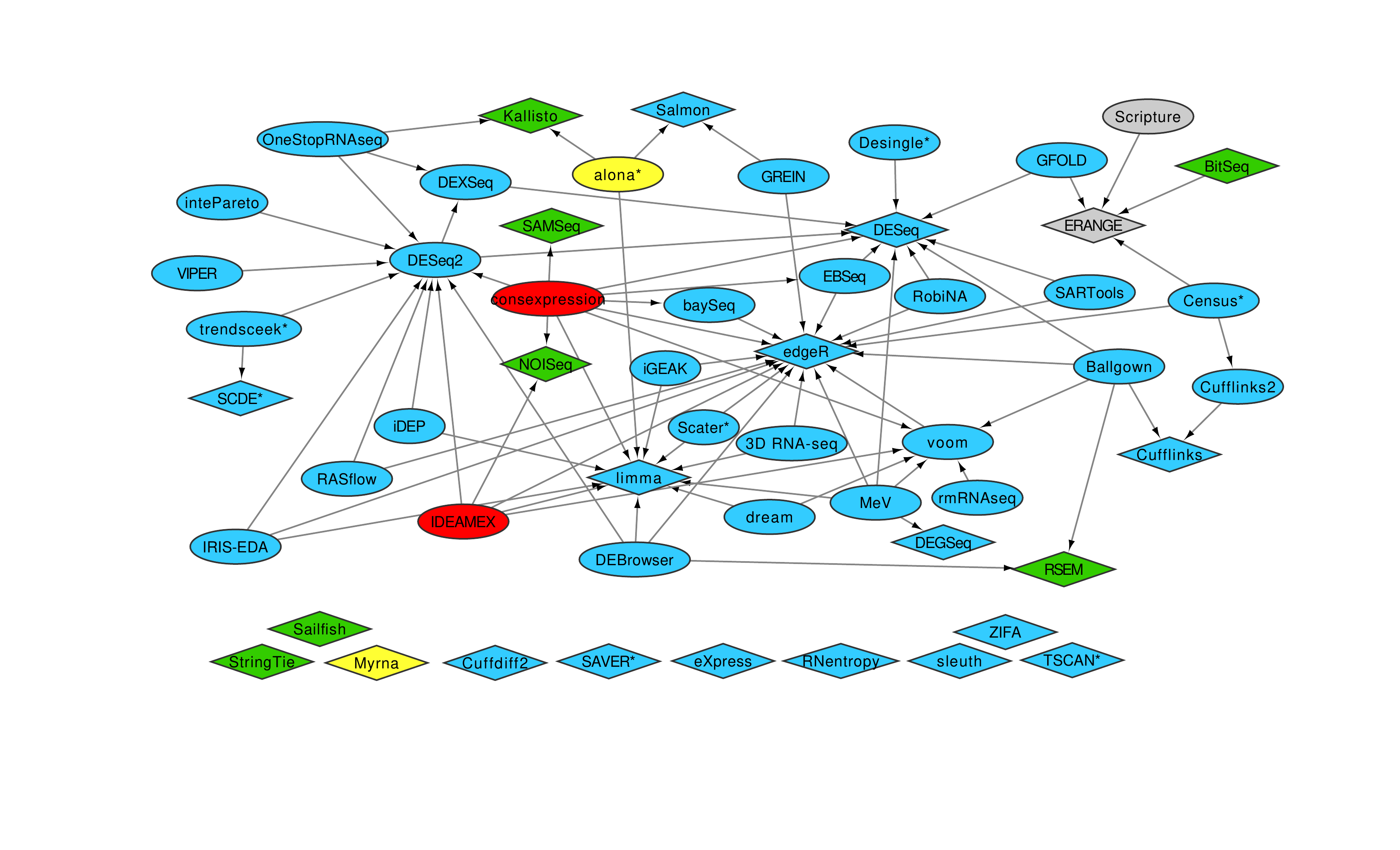}
    \caption{Network of interaction between methodologies for the analysis of differential gene expression. The edges represent that a tool uses partially or totally another one as a base, as an inheritance of the base method. The colours of the nodes show their method: blue (parametric), green (non-parametric), red (hybrid) and yellow (parametric or non-parametric). The diamond-shaped nodes show “seminal” studies (which do not depend on any other). The items that contain * in the name indicate methods developed in the context of single-cell sequencing analysis.}
    \label{fig:network}
\end{figure}

Regarding the methods for differential expression analysis, this review identified that many of the available computational tools dedicated to DEG analysis are derived from previous methods or use them as part of their solutions. Therefore, a study was performed to recover the relationships between computational tools from the current literature. As a result, a network of interactions between the computational tools for DEG analysis was produced, which is presented in the Figure \ref{fig:network}.

It is possible to notice that few methodologies can be considered totally original (diamond-shaped node in Figure \ref{fig:network}) which shows that the development of various computational methodologies for expression analysis are based on some specific methodologies.
It is possible to notice that edgeR \cite{robinson2010edger}, DESeq \cite{anders2010differential} and DESeq2 \cite{Love2014} and limma \cite{ritchie2015limma} methods are highlighted in this criterion, which are used by other methods as a basis for differential gene expression analysis.
More specifically, considering only the main methods available in the literature, edgeR is adopted by other 17 methods, DESeq and DESeq 2 by 10 and 9 methods respectively and, limma by ten methods.

Another interesting aspect that deserves to be highlighted is that the methodologies that adopt parametric statistical distributions in their analyses are more abundant in the literature (in blue) than the non-parametric (in green) and hybrid methodologies (in red). The discrepancy between the amount of parametric and non-parametric methodologies shows that there is a scenario that needs improvements in the methods, since the tools that use as the basis their predecessors bring some improvement or functionality to the DEG analysis.

No other tools were identified that associate results from parametric and non-parametric techniques to indicate DEG other than the consexpression tool \cite{Costa-Silva2017}. Therefore, it is a tool that addresses the gathering of methodologies, known as wisdom of crowds for decision making.

It is also possible to observe that some methodologies have not been adopted by other methods and appear at the bottom without edges in this network. The most recent methodologies dedicated to single-cell data analysis appear in yellow. Therefore, Figure \ref{fig:network} presents an overview of the relations considering the main methods identified in the current literature.

An issue that arises from the analysis of these interactions is that computational tools that implement non-parametric methods may represent a necessary solution, because, in the interaction network, the non-parametric methodologies are used as a basis for other parametric and hybrid methods. This may indicate a direction in the development of more adequate methods of analysis. 
In this scenario, there is a certain convergence between the parametric methods. 
Another point that draws attention is the opportunities to explore the development of non-parametric (data-driven) and hybrid methodologies.

\begin{figure}[!ht]
    \center
    \includegraphics[width=0.9\linewidth]{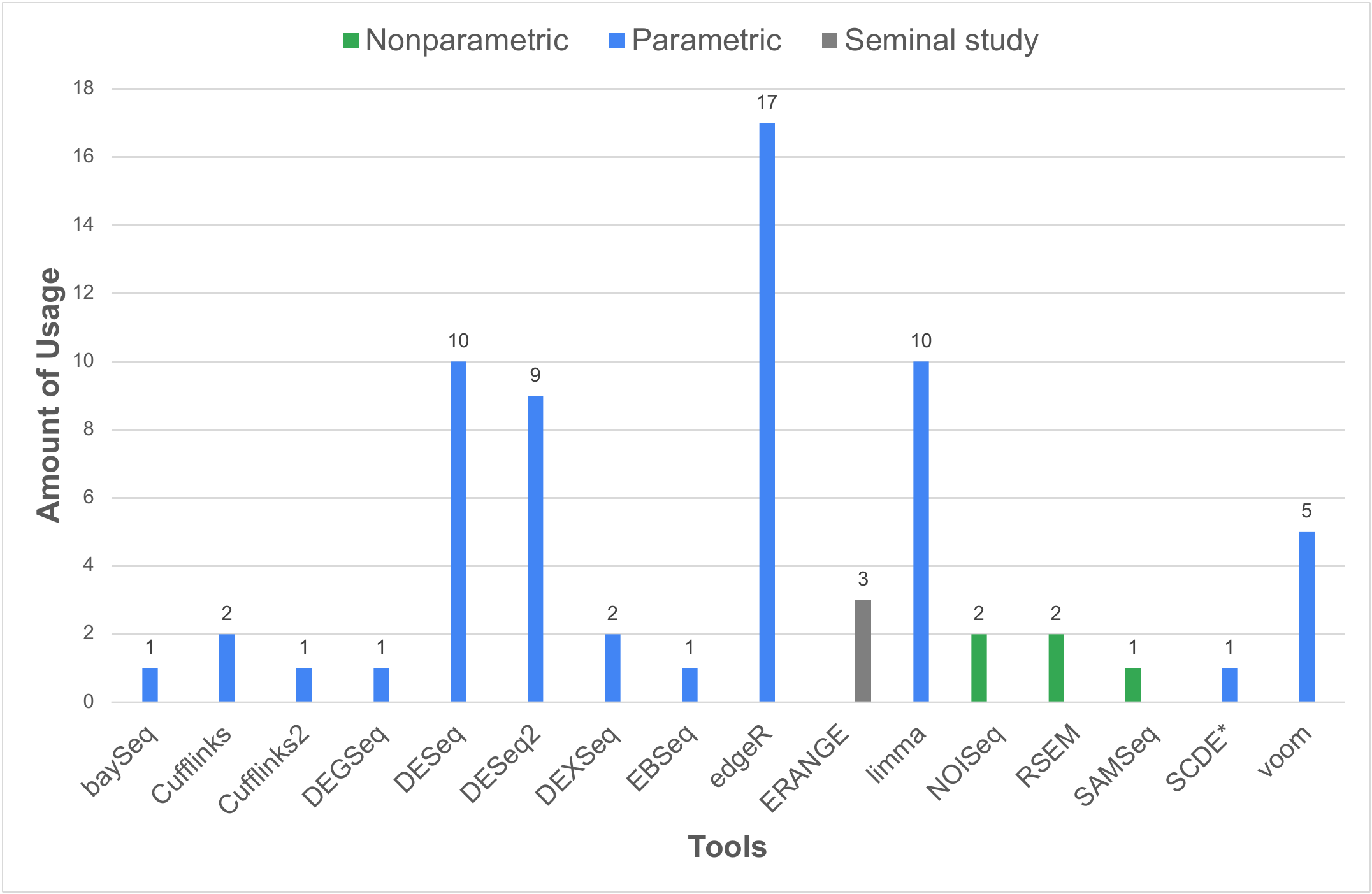}
    \caption{Histogram of the use of the methods, where the X axis presents only the tools used as a base (dependency) for some other. The bars are coloured according to the category of the tool, following the identification colours of the timeline (Figure \ref{fig:timeline}) and the interaction network (Figure \ref{fig:network}), where blue shows parametric methods, green non-parametric and grey ``seminal studies’’. The items that contain * in the name indicate tools developed in the context of single-cell sequencing analysis.}
    \label{fig:histogram}
\end{figure}

The observation regarding the number of use among of the tools considered in this review is also relevant, in order to point out the most used ones. The Figure \ref{fig:histogram} presents this analysis, indicating preference DESeq, DESeq2, edgeR and limma. 
The preference can be associated with maintenance, ease of use and the vast documentation made available by developers and community regarding these tools.

Among the 42 tools analyzed by this review, only 14 are used as a base for other studies, indicating that there are many tools but that most of them use a base method previously created. The data also show that most of the computational solutions for DEG analysis are based on the same analysis method.

Among the analyzed tools few methods were identified that do not use other methods as a base (in this review called original). Among them, approximately half are not used as a basis for the development of other methods, some because they were developed to be used in a defined pipeline, such as Cuffdiff2 \cite{trapnell2013differential} and sleuth \cite{Pimentel2017}. 
The full list of tools analysed in this review is available in supplementary material 1.

Single-cell RNA expression analysis (scRNA-seq or Single-cell) is revolutionising organismal science, allowing unbiased identification of previously uncharacterised molecular heterogeneity at the cellular level \cite{Pierson2015}.
Single-cell sequencing has become popular \cite{Breda2021} and, in this review, some tools for the expression analysis with single-cell data are pointed out with an ``*’’. 
In Figure~\ref{fig:network} it is possible to observe that the computational tools for single-cell analysis use as base tools used for RNA-Seq. The alona \cite{Franzen2020} tool uses limma \cite{ritchie2015limma}, while the DEsingle \cite{Miao2018} tool uses DESeq \cite{anders2010differential} and trendsceek \cite{Edsgard2018} uses DESeq2 \cite{Love2014}.

Although single-cell expression analysis is an important topic, it is still in its infancy and has its own characteristics. Clearly, there is a need that future studies can and should include analytical techniques for single-cell data, as an emerging topic and that deserves a dedicated review.


\section{Conclusion}
\label{sec:conclusion}

A review of computational methods for DEG analysis since the popularization of RNA-Seq in 2009 was presented, indicating and discussing the most popular tools in the current literature, bringing a contribution to the understanding of the steps involved, available methods, their particularities and applications.

The development of analysis pipelines by including new functionalities has become a trend. Avoiding the need for many replicates, required by RNA-Seq sequencing, and yet maintaining satisfactory results, is a challenge that deserves attention in the development of the DEG analysis methodologies.

The fundamental concepts and computational tools for the expression analysis, in which it is possible to identify the tendency to reuse methodologies in the development of computational tools (software), and also, to incorporate new functionalities to the existing software are presented.

Therefore, it was possible to verify that the context of parametric methodologies presents a more stable scenario, showing convergence with the methods available in the literature. In contrast, this review points out that there is a challenge in the development of non-parametric (data-driven) and hybrid methodologies for DEG analysis.

In conclusion, this review brings the discussion about different methodologies applied in differential expression analysis, contributes with notes and directions to the community, with the purpose of clarifying some aspects of the analysis and serve as support to data analysts in the context of bioinformatics.


 \section{Competing interests}
The authors declare that there is no conflict of interest.



\section{Acknowledgments}
INCT - Plant-Growth Promoting Microorganisms for Agricultural Sustainability and Environmental Responsibility (CNPq 465133/2014-4, Fundação Araucária-STI, CAPES). DSD research on DEG analysis are funded by CNPq (312823/2019-3) and FAPESP (2016/10896-0, 2018/08042-8 and 2019/15477-3).

\bibliographystyle{unsrt}
\bibliography{reference}
\end{document}